\documentclass{article}
\usepackage{amsmath}
\usepackage{eurosym}
\usepackage{graphicx}
\usepackage{float,cite}

\providecommand{\keywords}[1]{\textbf{Keywords:} #1}

\begin{document}

\title{A new non-commutative correction to the thermodynamics and evaporation of the Schwarzschild black hole in non-commutative gauge theory}
\author{Slimane Zaim and Fatma Zohra Bara\\
Department of Physics, Faculty of Matter Sciences,\\
University of Batna-1, Batna 05000, Algeria\\
zaim69slimane@yahoo.com}
\maketitle

\begin{abstract}
\sloppy
\hyphenpenalty=10000
\exhyphenpenalty=10000

We investigate the modified thermodynamic properties of a deformed Schwarzschild black hole (SBH) arising from new non-commutative (NC) corrections to the tetrad fields obtained through the Seiberg--Witten map. By deriving the corrected event horizon radius, we evaluate the corresponding thermodynamic quantities, including the Hawking temperature, entropy, heat capacity, Helmholtz free energy, and pressure. We also examine the evaporation process of the black hole within this non-commutative framework. The behavior of the heat capacity reveals a rich thermodynamic structure consisting of three distinct phases: two unstable phases with negative heat capacity, corresponding to a large black hole for $r_h > r_h^{\mathrm{c}}=\frac{3}{2}\Theta$ and a small black hole for $r_h < r_h^{(1)}=\frac{\sqrt{3}}{2}\Theta$, separated by an intermediate stable phase with positive heat capacity in the range $r_h^{(1)} < r_h < r_h^{\mathrm{c}}$. Our analysis also shows that the Schwarzschild black hole in non-commutative geometry possesses a finite lifetime that exceeds that of its classical counterpart, owing to the effective contribution of non-commutativity to the black hole mass. The NC correction naturally introduces a fundamental length scale,
$\Theta \approx 1.6\times10^{-35}\,\mathrm{m} \equiv l_{\mathrm{P}}$, which is of the order of the Planck length.
\end{abstract}


\section{Introduction}

Hawking showed that black holes are not completely dark, but instead emit thermal radiation with a blackbody spectrum as a consequence of quantum effects near the event horizon~\cite{1}. This remarkable discovery established a profound connection between gravity, quantum theory, and thermodynamics, allowing black holes to be treated as genuine thermodynamic systems characterized by well-defined temperature and entropy. Shortly thereafter, Bekenstein demonstrated that the entropy of a black hole is proportional to the area of its event horizon, providing a consistent thermodynamic interpretation of these objects~\cite{2}. This picture was further strengthened by Gibbons and Hawking, who derived the black hole temperature using quantum field theory in curved spacetime~\cite{3}. Although this semiclassical framework has been highly successful in describing black hole thermodynamics, it neglects the quantum nature of gravity. Consequently, its validity becomes questionable during the final stages of black hole evaporation, when the temperature increases rapidly as the black hole mass decreases~\cite{4}.

In recent years, considerable attention has been devoted to investigating the effects of quantum gravity on the thermodynamic properties of black holes, particularly Schwarzschild black holes. Various approaches have shown that quantum corrections, especially those arising from the generalized uncertainty principle, can significantly modify the Hawking temperature, heat capacity, stability conditions, and evaporation process. Likewise, quantum corrections to the black hole entropy within effective quantum gravity frameworks lead to deviations from the Bekenstein--Hawking area law and alter the thermodynamic stability and phase structure of these systems~\cite{5,6,7,8}. Despite these important advances, a complete and universally accepted theory of quantum gravity is still lacking. This situation motivates the study of non-commutative spacetime~\cite{9,10}, which provides an alternative geometric framework in which spacetime coordinates satisfy the canonical commutation relation
\begin{equation}\label{1}
[\hat{x}^{\mu},\,\hat{x}^{\nu}] = i\,\Theta^{\mu\nu},
\end{equation}
where $\Theta^{\mu\nu}$ is a constant antisymmetric tensor with dimensions of $(\mathrm{length})^{2}$. Throughout this work, we consider only space--space non-commutativity, namely $\Theta^{0i}=0$, in order to avoid the well-known problems with unitarity~\cite{11,12}. The non-commutative coordinate operators $\hat{x}^{\mu}$ are related to the ordinary coordinates through
\begin{equation}\label{2}
\hat{x}^{\mu}=x^{\mu}+\frac{1}{2}\Theta^{\mu\nu}\frac{\partial}{\partial x^{\nu}}.
\end{equation}

In non-commutative spacetime, the ordinary product of two functions $f(x)$ and $g(x)$ is replaced by the Moyal (star) product,
\begin{equation}\label{3}
(f\star g)(x)=f(x)\exp\left(\frac{i}{2}\Theta^{\mu\nu}\overleftarrow{\partial}_{\mu}\overrightarrow{\partial}_{\nu}\right)g(x).
\end{equation}
Within the framework of non-commutative gauge gravity, the Moyal product and the Seiberg--Witten (SW) map provide a systematic procedure for constructing gravitational theories on non-commutative spacetime. This formalism has been extensively applied to the study of black hole solutions~\cite{13,14,15}, including investigations of their thermodynamic properties and evaporation processes~\cite{16,17,18,19,20,21,22,23,24,25,26,27}. In particular, the recent work of Tajron \textit{et al.}~\cite{26} introduced new non-commutative corrections to the deformed tetrad fields $\hat{e}_{\mu}^{\,a}(r,\Theta)$ originally obtained in Ref.~\cite{27}. Motivated by these developments, we investigate the thermodynamic properties and evaporation process of the Schwarzschild black hole based on these newly corrected tetrad fields. Our objective is to determine how the modified non-commutative corrections affect the event horizon, thermodynamic quantities, stability, and evaporation of the Schwarzschild black hole within the framework of non-commutative gauge theory.

The remainder of this paper is organized as follows. In Section~2, we employ the newly corrected tetrad fields together with the Seiberg--Witten map to derive the non-commutative corrections to the Schwarzschild metric. In Section~3, we investigate the thermodynamic properties of the deformed Schwarzschild black hole by evaluating the Hawking temperature, entropy, heat capacity, Helmholtz free energy, and pressure up to second order in the non-commutative parameter $\Theta$. We also analyze the corresponding black hole evaporation process. Finally, in Section~4, we summarize our main results and present our conclusions. 

\section{Schwarzschild Black Holes in non-commutative gauge theory}

We consider the Schwarzschild black hole described by the static, spherically symmetric line element
\begin{equation}\label{4}
ds^{2}=-\left(1-\frac{2GM}{c^{2}r}\right)dt^{2}+\left(1-\frac{2GM}{c^{2}r}\right)^{-1}dr^{2}+r^{2}d\theta^{2}+r^{2}\sin^{2}\theta\,d\varphi^{2},
\end{equation}
where $c$, $G$, and $M$ denote the speed of light, the gravitational constant, and the black hole mass, respectively.

In the tetrad formulation of gravity based on the gauge group SO$(4,1)$, the spacetime metric $g_{\mu\nu}$ is related to the tetrad fields $e_{\mu}^{\,a}$ through
\begin{equation}
g_{\mu\nu}=\eta_{ab}e_{\mu}^{\,a}e_{\nu}^{\,b},\label{5}
\end{equation}
where $\eta_{ab}$ is the Minkowski metric. A convenient diagonal tetrad reproducing the Schwarzschild metric~\eqref{4} is
\begin{equation}
e_{\mu}^{\,a}=\left(
\begin{array}{cccc}
\sqrt{1-\frac{2GM}{c^{2}r}} & 0 & 0 & 0\\
0 & \displaystyle\frac{1}{\sqrt{1-\frac{2GM}{c^{2}r}}} & 0 & 0\\
0 & 0 & r & 0\\
0 & 0 & 0 & r\sin\theta
\end{array}
\right).
\label{6}
\end{equation}

Within non-commutative gauge gravity, the deformed tetrad fields $\hat{e}_{\mu}^{\,a}(r,\Theta)$ can be expanded perturbatively in powers of the non-commutative parameter $\Theta$ using the SW map~\cite{28},
\begin{equation}\label{7}
\hat{e}_{\mu}^{\,a}(r,\Theta)=e_{\mu}^{\,a}(r)-i\,\Theta^{\nu\rho}e_{\mu\nu\rho}^{\,a}(r)+\Theta^{\nu\rho}\Theta^{\lambda\tau}e_{\mu\nu\rho\lambda\tau}^{\,a}(r)+\mathcal{O}(\Theta^{3}),
\end{equation}
where
\begin{align}
e_{\mu\nu\rho}^{\,a}&=\frac{1}{4}\left[\omega_{\nu}^{ac}\,\partial_{\rho}e_{\mu }^{\,d}+\left(\partial_{\rho}\omega_{\mu}^{ac}+R_{\rho\mu}^{ac}\right)e_{\nu}^{\,d}\right]\eta_{cd},\label{8}\\
e_{\mu\nu\rho\lambda\tau}^{\,a}&=\frac{1}{32}\left[2\{F_{\tau\nu},F_{\mu\rho}\}^{ab}e_{\lambda }^{\,c}-\omega_{\lambda }^{ab}(D_{\rho}F_{\tau\nu}^{cd}+\partial_{\rho}F_{\tau\nu}^{cd})e_{\nu}^{\,m}\eta_{dm}\right.\notag\\
&\left.-\{\omega_{\nu},(D_{\rho}F_{\tau\nu}+\partial_{\rho}F_{\tau\nu})\}^{ab}e_{\lambda}^{\,c}-\partial_{\tau}\{\omega_{\nu},(\partial_{\rho}\omega_{\mu}+F_{\rho\mu})\}^{ab}e_{\lambda}^{\,c}\right.\notag\\
&\left.-\omega_{\lambda}^{ab}\left(\omega_{\nu}^{cd}\partial_{\rho}e_{\mu}^{\,m}+\left(\partial_{\rho}\omega_{\mu}^{cd}+F_{\rho\mu}^{cd}\right)e_{\nu}^{\,m}\right)\eta_{dm}+2\partial_{\nu}\omega_{\lambda}^{ab}\partial_{\rho}\partial_{\tau}e_{\mu}^{\,c} \right.\notag\\
&\left.-2\partial_{\rho}\left(\partial_{\tau}\omega_{\mu}^{ab}+F_{\tau\mu}^{ab}\right)\partial_{\nu}e_{\lambda}^{\,c}-\{\omega_{\nu},(\partial_{\rho}\omega_{\lambda}+F_{\rho\lambda})\}^{ab}\partial_{\tau}e_{\mu}^{\,c}\right.\notag\\
&\left.-\left(\partial_{\tau}\omega_{\mu}+F_{\tau\mu}\right)\left(\omega_{\nu}^{cd}\partial_{\rho}e_{\lambda}^{\,m}+\left((\partial_{\rho}\omega_{\lambda}+F_{\rho\lambda})\right)e_{\nu}^{\,m}\right)\eta_{dm}\right]\eta_{cb}\notag\\
&-\frac{1}{16}\omega_{\lambda}^{ac}\omega_{\nu}^{db}e_{\rho}^{\,f}R_{\tau\mu}^{gm}\eta_{cd}\eta_{fg}\eta_{bm}, \label{9}
\end{align}
and 
\begin{align}
\{\alpha,\beta\}^{ab}=&\left(\alpha^{ac}\beta^{db}+\beta^{ac}\alpha^{db}\right)\eta_{cd},\qquad[\alpha,\beta]^{ab}=\left(\alpha^{ac}\beta^{db}-\beta^{ac}\alpha^{db}\right)\eta_{cd},\notag\\
& D_{\mu }R_{\rho\sigma}^{ab}=\partial_{\mu}R_{\rho\sigma}^{ab}+\left(\omega_{\mu}^{ac}R_{\rho\sigma}^{db}+\omega_{\mu}^{bc}R_{\rho\sigma}^{da}\right)\eta_{cd}.\label{10}
\end{align}

The deformed spacetime metric is constructed from the non-commutative tetrad fields according to~\cite{15}
\begin{equation}
\hat{g}_{\mu\nu}(r,\Theta)=\frac{1}{2}\left(\hat{e}_{\mu}^{\,a}\star\hat{e}_{\nu}^{\,b\dagger}+\hat{e}_{\nu}^{\,b}\star\hat{e}_{\mu}^{\,a\dagger}\right)\eta_{ab},
\label{11}
\end{equation}
where $\hat{e}_{\mu}^{\,a\dagger}$ denotes the Hermitian conjugate of the deformed tetrad field. From Eq.~\eqref{7}, one obtains
\begin{equation}\label{12}
\hat{e}_{\mu}^{\,a\dagger}(r,\Theta)=e_{\mu}^{\,a}(r)+i\,\Theta^{\nu\rho}e_{\mu\nu\rho}^{\,a}(r)+\Theta^{\nu\rho}\Theta^{\lambda\tau}e_{\mu\nu\rho\lambda\tau}^{\,a}(r)+\mathcal{O}(\Theta^{3}).
\end{equation}

Throughout this work, we assume that the only non-vanishing component of the non-commutativity tensor is
\begin{equation}\label{13}
\Theta^{12}=\Theta=-\Theta^{21},\qquad\mu,\nu=0,1,2,3,
\end{equation}
where $\Theta$ is a positive real parameter that characterizes the strength of spacetime non-commutativity.

Following the treatment of Ref.~\cite{26}, which revises the non-commutative correction term for the tetrad fields $\hat{e}_{\mu }^{\,a}(r,\Theta )$ originally derived in Ref.~\cite{27}, the non-vanishing components of the deformed tetrad fields for the Schwarzschild black hole solution, calculated up to second order in $\Theta$, given in Eq. \eqref{8}, are
\begin{align}
\hat{e}_{t}^{0}&=\left(1-\frac{2M}{r}\right)^{1/2}+\frac{\Theta^{2}}{2r^{4}}\,M\left(11\,M-4\,r\right)+\mathcal{O}(\Theta^{3}),\label{14}\\
\hat{e}_{r}^{1}&=\left(1-\frac{2M}{r}\right)^{-1/2}-\frac{\Theta^{2}}{4\,r^{4}}\,M\left(2\,r-3\,M\right)\left(1-\frac{2M}{r}\right)^{-3/2}+\mathcal{O}(\Theta^{3}),\label{15}\\
\hat{e}_{\theta }^{1}&=-\frac{i}{4}\left(1-\frac{2M}{r}\right)^{-1/2}\Theta+\mathcal{O}(\Theta^{2}),\label{16}\\
\hat{e}_{\theta }^{2}& =r-\frac{M}{r^{2}}\Theta^{2}+\mathcal{O}(\Theta^{3}),\label{17}\\
\hat{e}_{\varphi }^{3}& =r\sin\theta-\frac{i}{4}\Theta\cos\theta -\frac{\Theta^{2}}{4r^{2}}M\left(r-M\,\right)\left(r-2M\right)^{-1}\sin\theta+\mathcal{O}(\Theta ^{3}).\label{18}
\end{align}
Substituting Eqs.~\eqref{14}--\eqref{18} into Eq.~\eqref{11}, we obtain the corresponding non-commutative corrections to the Schwarzschild metric,
\begin{align}
-\hat{g}_{tt}&=\left(1-\frac{2M}{r}\right)+\frac{\Theta^{2}}{2r^{4}}\,M\left(11\,M-4\,r\right)+\mathcal{O}(\Theta^{3}),\label{19}\\
\hat{g}_{rr}&=\left(1-\frac{2M}{r}\right)^{-1}-\frac{\Theta^{2}}{2\,r^{2}}\,M\left(2\,r-3\,M\right)\left(r-2M\right)^{-2}+\mathcal{O}(\Theta^{3}),\label{20}\\
\hat{g}_{\theta\theta}& =r^{2}+\frac{\Theta^{2}}{16r}\left(64\,M^{2}-32\,M\,r+r^{2}\right)\left(r-2M\right)^{-1}+\mathcal{O}(\Theta^{3}),\label{21}\\
\hat{g}_{\varphi\varphi}& =r^{2}\sin^{2}\theta+\frac{\Theta^{2}}{4}\left(\frac{5}{4}\cot^{2}\theta+\frac{\left(2\,M^{2}-4\,M\,r+r^{2}\right)}{r\left(r-2M\right)}\right)\sin^{2}\theta+\mathcal{O}(\Theta^{3}).\label{22}
\end{align}

In the commutative limit $\Theta\rightarrow0$, the obtained metric reduces to the standard Schwarzschild solution. Although the corrected metric remains diagonal, the non-commutative deformation breaks the spherical symmetry of the spacetime. The deformed metric $\hat{g}_{\mu\nu}(r,\Theta)$ possesses the usual curvature singularity at $r=0$ together with a horizon determined by the condition $1/\hat{g}_{rr}=0$.
To leading order in $\Theta$, the inverse radial metric component is
\begin{align}
1/\hat{g}_{rr}&=\left(1-\frac{2M}{r}\right)\left[1+\frac{\Theta^{2}}{2\,r^{4}}\,\frac{M\left(3\,M-2\,r\right)}{\left(1-\frac{2M}{r}\right)}\right]^{-1}\notag\\
&\approx 1-\frac{2M}{r}-\frac{M\left(3\,M-2\,r\right)}{2\,r^{4}}\Theta^{2}+\mathcal{O}(\Theta^{3}),\label{23}
\end{align}

It is worth noting that, at $r=2M$, Eq.~\eqref{23} yields the finite value
\[\hat{g}_{rr}^{-1}=\frac{\Theta^{2}}{32M^{2}},\]
showing that $r=2M$ is no longer the location of the event horizon once non-commutative corrections are included. To determine the corrected event horizon analytically, we introduce the perturbative ansatz~\cite{9}
\begin{equation}
r_{h}^{\mathrm{NC}}=r_{h}+\delta\Theta^{2},\label{24}
\end{equation}
and retain terms up to $\mathcal{O}(\Theta^{2})$. Expanding Eq.~\eqref{23} then gives
\begin{equation}
\delta=-\frac{1}{16M},\label{25}
\end{equation}
from which the non-commutative event horizon is found to be
\begin{equation}
r_{h}^{\mathrm{NC}}=r_{h}\left[1-\frac{1}{8}\left(\frac{\Theta}{r_{h}}\right)^{2}\right],\label{26}
\end{equation}
where $r_{h}=2M$ is the Schwarzschild event horizon in the commutative limit ($\Theta=0$).

We emphasize that Ref.~\cite{26} interpreted the finite value
\[\hat{g}_{rr}^{-1}(r=2M)=\frac{\Theta^{2}}{32M^{2}}\]
as implying that the event horizon remains unchanged. However, since the horizon is defined by the condition $\hat{g}_{rr}^{-1}=0$, the radius $r=2M$ no longer satisfies the horizon equation once non-commutative corrections are included. Consequently, the event horizon is shifted according to Eq.~\eqref{26}. Interestingly, the resulting expression agrees with the corrected event horizon obtained in Ref.~\cite{29}. 

\section{Thermodynamic properties}

The mass of the Schwarzschild black hole in the framework of non-commutative gauge theory of gravity is obtained, up to second order in the non-commutative parameter $\Theta$, by solving the horizon equation
\[\left.\hat{g}_{tt}(r,\Theta)\right|_{r=r_h^{NC}}=0\]
with respect to the mass parameter $M$. The resulting expression is
\begin{equation}\label{27}
\hat{M}=\frac{r_h}{2}+\frac{\Theta^{2}}{8r_h},
\end{equation}
where $r_h$ denotes the Schwarzschild horizon radius.

In the commutative limit ($\Theta\rightarrow0$), Eq.~(\ref{27}) reduces to the standard Schwarzschild mass, $\hat{M}=r_h/2$. The second term represents the correction induced by spacetime non-commutativity. This correction coincides with that reported in Ref.~\cite{22}, although it differs from the corresponding result obtained in Ref.~\cite{25}. Equation~(\ref{27}) further shows that, for $\Theta\neq0$, the black-hole mass diverges as $r_h\rightarrow0$, reflecting the persistence of the central singularity. Moreover, the corrected mass possesses a minimum value, $\hat{M}_{\rm min}=\frac{\Theta}{2}$, which occurs at $r_h=r^0=\frac{\Theta}{2}$. This behavior indicates that non-commutative effects acts like the gravitational force and is consistent with the results reported in Ref.~\cite{22}.
\begin{figure}[H]
    \centering
    \includegraphics[width=0.9\linewidth]{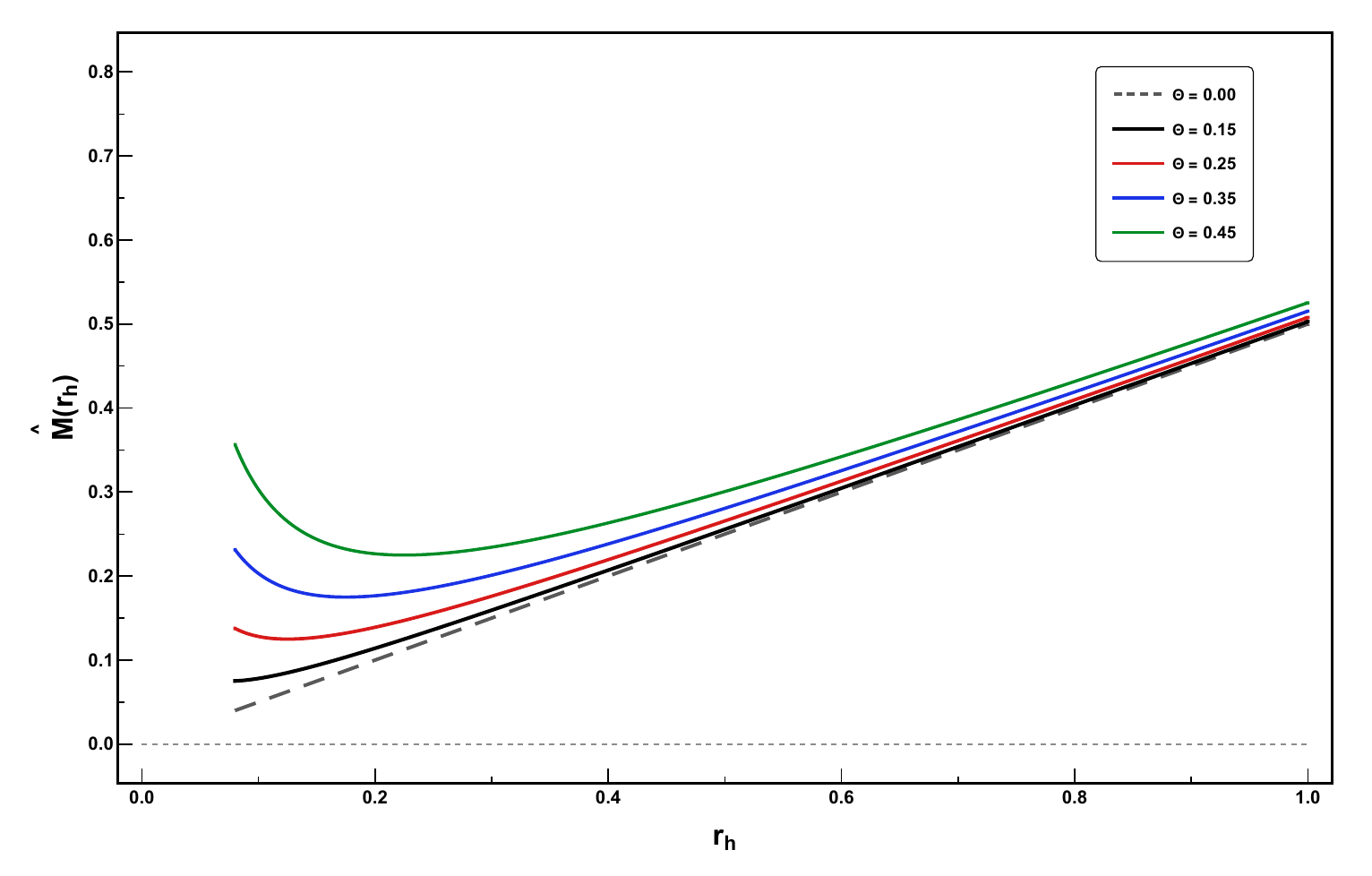}
    \caption{Corrected mass $\hat{M}(r_h)$ of the non-commutative Schwarzschild black hole for different values of the non-commutative parameter $\Theta$.\label{fig:Mhat_rh}}
    
\end{figure}

\subsection{Hawking temperature}

We now investigate the thermodynamic properties of the non-commutative Schwarzschild black hole and its phase transition, with particular emphasis on its Hawking temperature and phase structure. Following the standard semiclassical approach based on the surface gravity of the event horizon \cite{1,2}, the Hawking temperature in non-commutative spacetime is defined as
\begin{equation}\label{28}
\hat{T}_{H}=\frac{\hat{\kappa}}{2\pi}=-\frac{1}{4\pi}\left.\frac{d\hat{g}_{tt}}{dr}\right|_{r=r_h^{NC}},
\end{equation}
where the derivative of the deformed metric component is
\begin{equation}\label{29}
\left.\frac{d\hat{g}_{tt}}{dr}\right|_{r=r_h^{NC}}=-\frac{2M}{r_h^2}+\frac{3\Theta^2}{4r_h^3}.
\end{equation}

Expanding Eq.~\eqref{28} perturbatively up to second order in $\Theta$ and retaining terms up to $\mathcal{O}(1/r_h^4)$ yields the analytical expression
\begin{equation}\label{30}
\hat{T}_{H}(r_h)=\frac{1}{4\pi r_h}-\frac{9\Theta^2}{16\pi r_h^3}.
\end{equation}

Equation~\eqref{30} shows that the Hawking temperature depends explicitly on both the non-commutative parameter $\Theta$ and the classical Schwarzschild horizon radius $r_h$. In the commutative limit, $\Theta\rightarrow0$, the standard Hawking temperature is recovered, $T_H=\frac{1}{4\pi r_h}$.

\begin{figure}[H]
    \centering
    \includegraphics[width=0.9\linewidth]{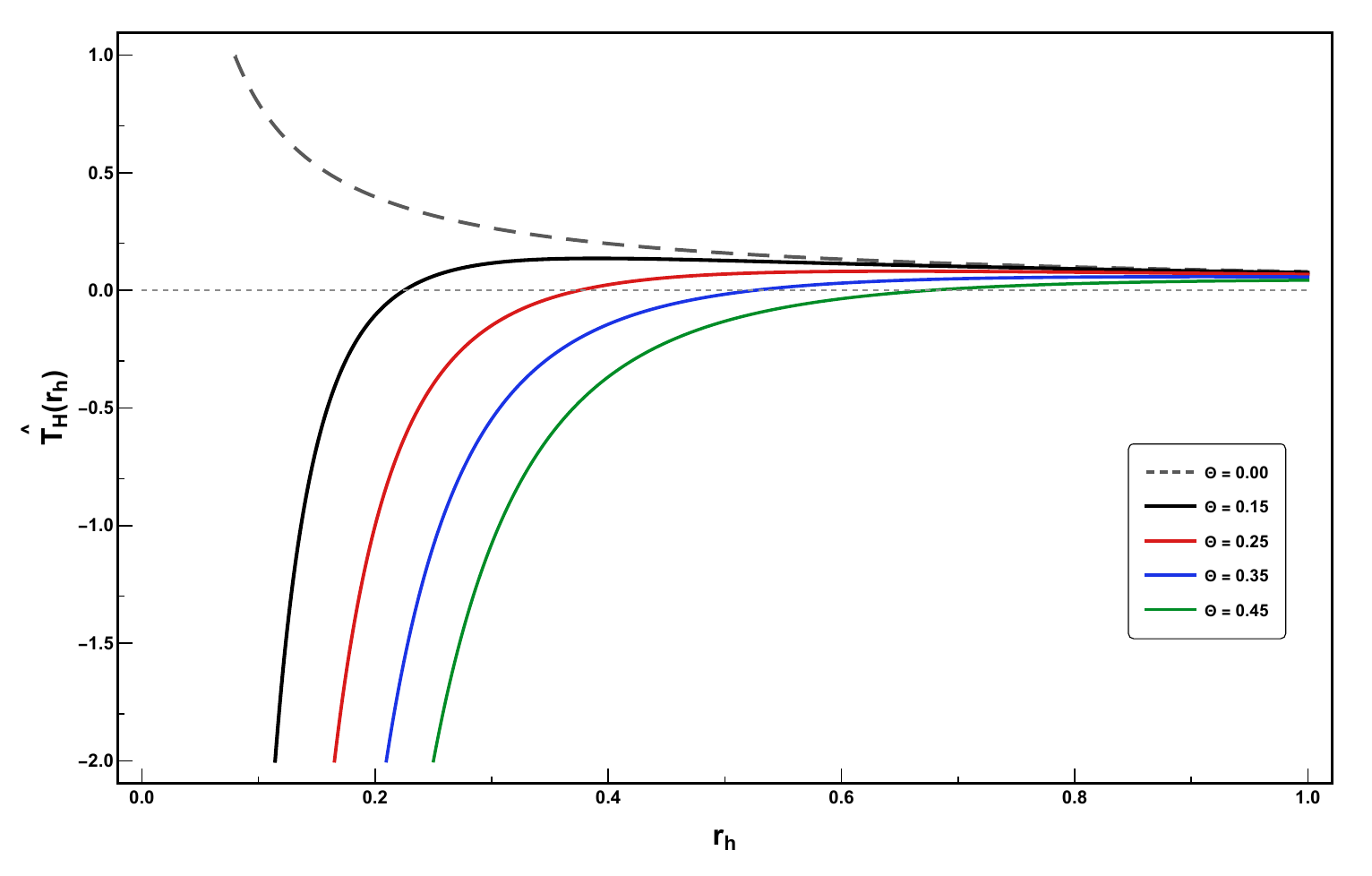}
    \caption{Corrected Hawking temperature $\hat{T}_{H}(r_h)$ of the non-commutative Schwarzschild black hole for different values of the non-commutative parameter $\Theta$.}
    \label{fig:That_rh}
\end{figure}

Figure~\ref{fig:That_rh} illustrates the behavior of the corrected Hawking temperature $\hat{T}(r_h)$ for different values of $\Theta$. In the commutative case ($\Theta=0$), the temperature diverges as $r_h\rightarrow0$, reproducing the well-known Schwarzschild result. In contrast, non-commutative effects remove this divergence. The corrected temperature reaches a finite maximum, $\hat{T}_H^{\rm max}\simeq 0.02/\Theta$, at the critical horizon radius $r_h^c\simeq2.6\,\Theta$, after which it decreases and eventually vanishes as the horizon radius approaches its minimum value, $r_h\simeq1.5\,\Theta$. As the non-commutative parameter $\Theta$ decreases, the maximum temperature increases. The existence of a maximum temperature, $\hat{T}^{\max}$, signals a thermodynamic phase transition, since the corresponding heat capacity diverges at this critical point. Compared with the results of Ref.~\cite{22}, these predictions are improved owing to the corrected non-commutative contributions to the metric components obtained in Ref.~\cite{26}.

The temperature profile can also be used to estimate the magnitude of the non-commutative parameter. At the onset of back-reaction, the thermal energy is assumed to be of the order of the maximum Hawking temperature $\hat{T}^{\max}$. In natural units ($\hbar=k_B=c=1$), $E_{\mathrm{th}}\simeq\hat{T}_H^{\rm max}$. At the corresponding critical radius $r_h^c$, the black-hole mass is
\begin{equation}
M=\frac{r_h^c}{2G}+\frac{\Theta^2}{8Gr_h^c}\simeq1.34\,\Theta\,M_{\rm Pl}^2,
\end{equation}
where $M_{\rm Pl}$ denotes the Planck mass.

Requiring the thermal energy to be comparable to the black-hole mass at the critical point provides an estimate for the non-commutative parameter,
\begin{equation}\label{31}
\Theta\approx 1.6\times10^{-35}\,\mathrm{m} \equiv l_{\mathrm{P}},
\end{equation}
indicating that the characteristic scale of spacetime non-commutativity is of the order of the Planck length.

\subsection{Entropy}

Several approaches have been developed to derive the entropy of black holes \cite{30,31}. Among the most widely used are the first law of black-hole mechanics, which relates the entropy to the area of the event horizon, and the first law of thermodynamics, which treats the black hole as a thermodynamic system. In the present work, we determine the entropy by integrating the first law of thermodynamics,
\begin{equation}
d\hat{M}=\hat{T}_{H}\,d\hat{S},
\label{32}
\end{equation}
where $\hat{M}$, $\hat{T}_H$, and $\hat{S}$ denote the non-commutative mass, Hawking temperature, and entropy, respectively.

Using Eqs.~\eqref{27}, \eqref{30}, the entropy is obtained as

\begin{align}
\hat{S}&=\int \frac{d\hat{M}}{\hat{T}_{H}}=\pi r_h^2+\frac{1}{2}\pi\Theta^2\ln\left(\pi r_h^2\right)\notag\\
&=S_{\rm SBH}+\frac{1}{2}\pi\Theta^2\ln S_{\rm SBH},\label{33}
\end{align}
where $S_{\rm SBH}=\pi r_h^2$ is the standard Bekenstein--Hawking entropy of the Schwarzschild black hole.

It is noteworthy that the entropy given by Eq.~(\ref{33}) has exactly the same functional form as the logarithmically corrected quantum entropy (in Planck units),
\begin{equation}\label{34}
S_q=\pi r_h^2+\alpha\ln\left(\pi r_h^2\right)+\text{constant},
\end{equation}
where the coefficient $\alpha$ depends on the underlying quantum theory of gravity \cite{33,34,35,36}. In loop quantum gravity, one typically finds a negative coefficient, $\alpha=-1/2$, whereas in string theory $\alpha$ is related to the four-dimensional central charge and may be either positive or negative, depending on the field content.

Within the present non-commutative gauge theory of gravity, the logarithmic correction is characterized by the positive coefficient $\alpha=\pi\Theta^2/2l_P^2$.

Using the estimate of the non-commutative parameter obtained in Eq.~\eqref{31}, namely $\Theta\simeq l_P$, one finds $\alpha\simeq\pi/2$.

Although the magnitude of this coefficient is comparable to those obtained in previous studies \cite{33,34}, its sign is opposite to that predicted by loop quantum gravity. Consequently, the non-commutative gauge theory considered here predicts a different behavior for black-hole evaporation, leading to a suppression of the Hawking emission rate in the final stage of evaporation, in contrast to the to loop quantum gravity scenario discussed in Ref.~\cite{29}.

\begin{figure}[H]
    \centering
    \includegraphics[width=0.9\linewidth]{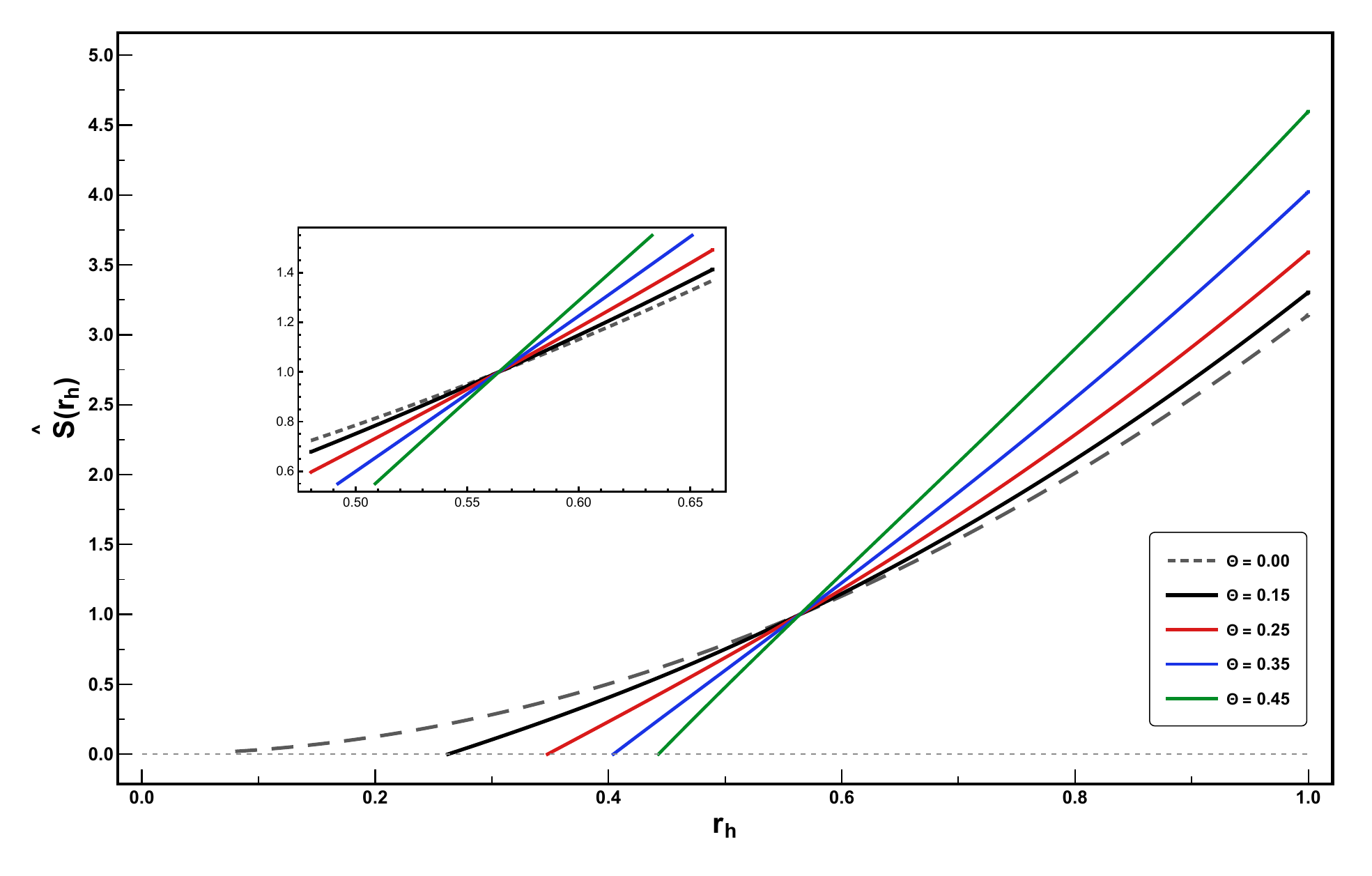}
    \caption{The corrected entropy $\hat{S}(r_h)$ of the non-commutative Schwarzschild black hole for different values of the parameter $\Theta$.}
    \label{fig:Shat_rh}
\end{figure}

Figure~\ref{fig:Shat_rh} shows the corrected entropy as a function of the horizon radius for several values of $\Theta$. As the black hole approaches the final stage of evaporation, the entropy decreases monotonically and eventually vanishes at the minimum horizon radius. This behavior indicates the existence of a non-zero minimum black-hole mass, corresponding to a stable remnant.

\subsection{Heat capacity}

To investigate the thermodynamic stability of the non-commutative Schwarzschild black hole, we analyze its heat capacity, which is defined by
\begin{align}
\hat{C}&=\hat{T}_{H}\left(\frac{\partial \hat{S}}{\partial \hat{T}_{H}}\right)=\hat{T}_{H}\left(\frac{\partial \hat{S}}{\partial r_h}\right)\left(\frac{\partial \hat{T}_{H}}{\partial r_h}\right)^{-1}\notag\\
&=-\pi\frac{\left(2r_h^{2}+\Theta^{2}\right)\left(4r_h^{2}-3\Theta^{2}\right)}{4r_h^{2}-9\Theta^{2}},\label{35}
\end{align}
where the final expression is obtained using the Hawking temperature given by Eq.~\eqref{30} together with the entropy in Eq.~\eqref{33}.

\begin{figure}[H]
    \centering
    \includegraphics[width=0.9\linewidth]{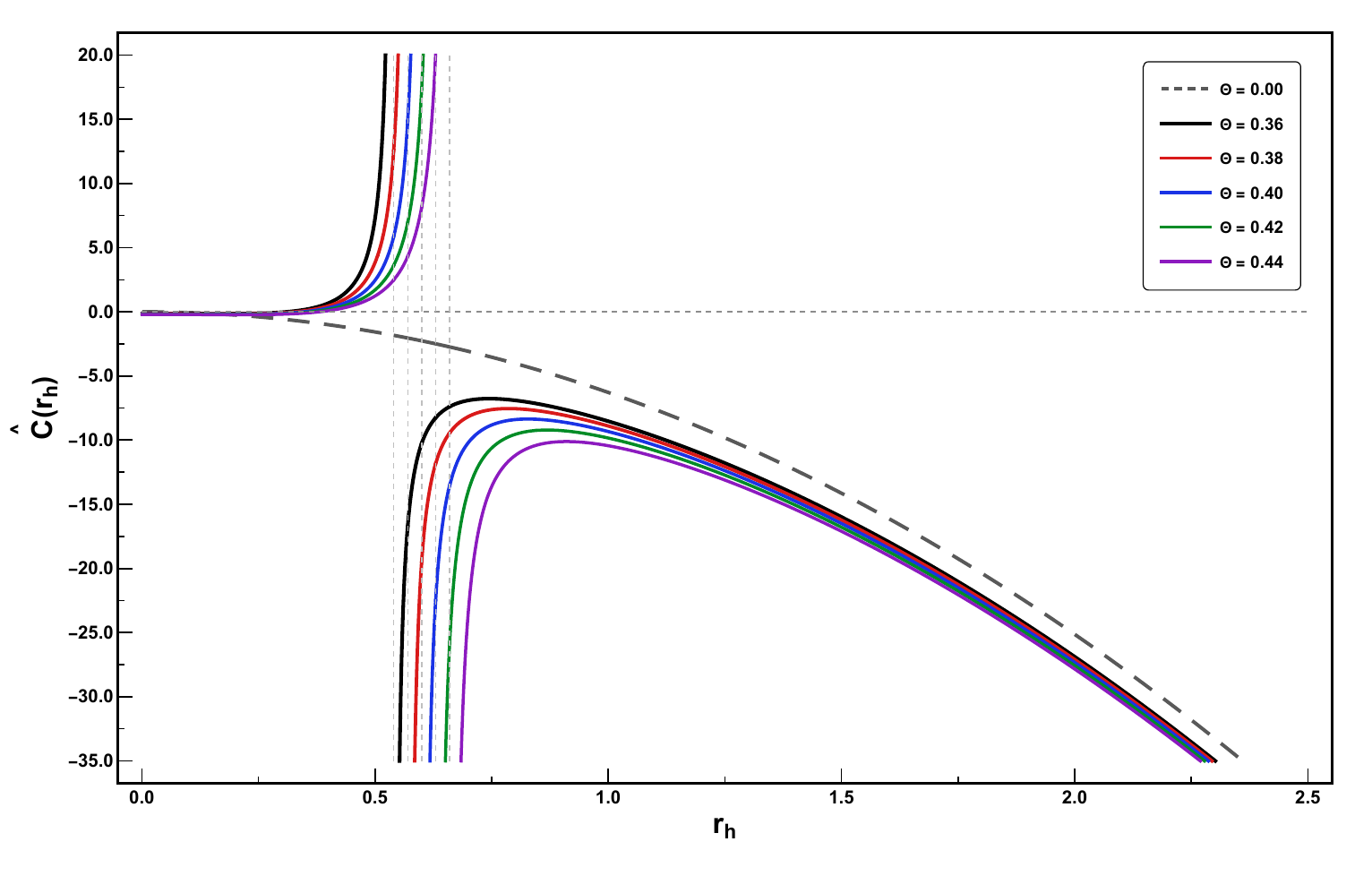}    
    \caption{Corrected heat capacity $\hat{C}(r_h)$ of the non-commutative Schwarzschild black hole for different values of the non-commutative parameter $\Theta$.}
    \label{fig:Chat_rh}
\end{figure}

Figure~\ref{fig:Chat_rh} illustrates the behavior of the corrected heat capacity $\hat{C}(r_h)$ as a function of the event-horizon radius for several values of the non-commutative parameter $\Theta$.

In black-hole thermodynamics, the sign of the heat capacity determines the local thermodynamic stability of the system. A positive heat capacity corresponds to a locally stable configuration, whereas a negative heat capacity indicates thermodynamic instability. The latter is the well-known behavior of the classical Schwarzschild black hole. Furthermore, divergences of the heat capacity are generally interpreted as signatures of second-order phase transitions.

These divergences signal critical behavior, indicating points at which the thermodynamic stability of the black hole changes and a phase transition may occur. When the heat capacity vanishes, the black-hole mass becomes insensitive to variations in its temperature. At this stage, the Hawking radiation ceases, the evaporation process terminates, and the black hole reaches a remnant state with constant mass.

From the expression for $\hat{C}$ in Eq.~(\ref{35}), we find that the heat capacity diverges at the critical horizon radius $r_h^{c}=\frac{3}{2}\Theta$, while it vanishes at $r_h^{(1)}=\frac{\sqrt{3}}{2}\Theta$. For a fixed value of $\Theta$, the heat capacity satisfies $\hat{C}>0$ for $r_h^{(1)}<r_h<r_h^{c}$, whereas $\hat{C}<0$ for $r_h<r_h^{(1)}$ and $r_h>r_h^{c}$. Therefore, the black hole undergoes changes in its thermodynamic stability at the critical radii $r_h^{(1)}$ and $r_h^{c}$, as illustrated in Fig.~\ref{fig:Chat_rh}. The results indicate that both small and large black holes possess negative heat capacity and are therefore thermodynamically unstable, whereas an intermediate class of black holes, satisfying $r_h^{(1)}<r_h<r_h^{c}$, has positive heat capacity, corresponding to a locally thermodynamically stable phase.

The divergence of the heat capacity signals the occurrence of a phase transition in the black hole. On the other hand, the condition $\hat{C}=0$ is reached at $r_h^{(1)}=\frac{\sqrt{3}}{2}\,\Theta$, which corresponds to a minimum remnant mass, $M_{\rm r}^{\rm min}=\sqrt{\frac{3}{8}}\,\Theta$. Since the remnant mass is proportional to the non-commutative parameter $\Theta$, the black hole cannot evaporate completely. Instead, the evaporation process terminates with a finite remnant, suggesting that some information may be preserved in this stable final state. Such black-hole remnants could also serve as primordial seeds for black-hole formation, subsequently growing through accretion and mergers to evolve into the supermassive black holes observed today.

\subsection{Helmholtz free energy}

In non-commutative spacetime, the Helmholtz free energy is defined by
\begin{equation}
\hat{F}=\hat{M}-\hat{T}_{H}\hat{S},\label{36}
\end{equation}
where $\hat{M}$, $\hat{T}_{H}$, and $\hat{S}$ denote the corrected mass, Hawking temperature, and entropy, respectively.

Substituting Eqs.~\eqref{27}, \eqref{30}, and \eqref{33} into Eq.~\eqref{36}, the Helmholtz free energy takes the form
\begin{equation}
\hat{F}=\frac{r_h}{4}+\frac{5\Theta^2}{16r_h}\left(1-\frac{2}{5}\ln\left(\pi r_h^2\right)\right).
\label{37}
\end{equation}

\begin{figure}[H]
\centering
\includegraphics[width=0.9\linewidth]{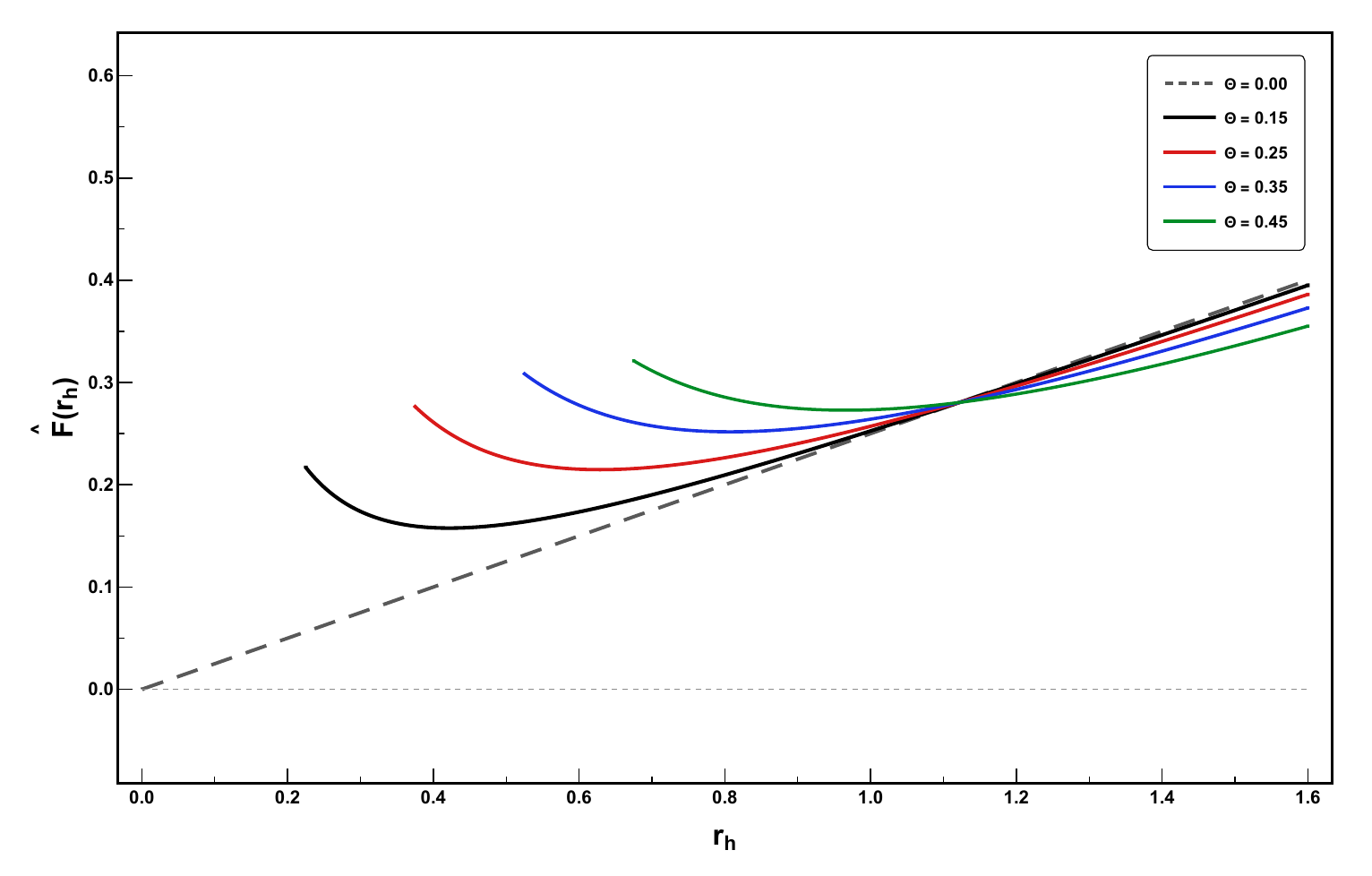}
\caption{Corrected Helmholtz free energy $\hat{F}(r_h)$ of the non-commutative Schwarzschild black hole for different values of the non-commutative parameter $\Theta$.}
\label{fig:Fhat_rh}
\end{figure}

Figure~\ref{fig:Fhat_rh} illustrates the behavior of the corrected Helmholtz free energy as a function of the horizon radius for several values of the non-commutative parameter $\Theta$. The free energy exhibits a well-defined minimum, indicating the existence of thermodynamic stability of the black hole. For example, for $\Theta=1$, the minimum occurs at approximately $r_h\simeq1.7$,  with $\hat{F}_{\rm min}\simeq0.25$. Moreover, the location of the minimum shifts toward larger values of the horizon radius as the non-commutative parameter $\Theta$ increases.

For the infinitesimal deformation considered in the present work, the event-horizon radius and the black-hole volume remain related through the usual geometric expression. Using Eq.~\eqref{26}, the corrected thermodynamic volume is
\begin{equation}
\hat{V}=\frac{4}{3}\pi\left(r_h^{NC}\right)^3\simeq\frac{4}{3}\pi r_h^3\left[1-\frac{3}{8}\left(\frac{\Theta}{r_h}\right)^2\right].\label{38}
\end{equation}

Treating the black hole as a thermodynamic system to deduce the pressure of in non-commutative space as
\begin{equation}
\hat{P}=\frac{\partial\hat{M}}{\partial\hat{V}}=\frac{\partial\hat{M}/\partial r_h}{\partial\hat{V}/\partial r_h}.\label{39}
\end{equation}
Differentiating Eqs.~\eqref{27} and \eqref{38} with respect to $r_h$ and retaining terms up to second order in $\Theta$, one obtains
\begin{equation}
\hat{P}=-\frac{1}{8\pi r_h^2}+\frac{3}{64\pi r_h^4}\Theta^2.\label{40}
\end{equation}

The first term represents the classical Schwarzschild contribution, whereas the second term is the correction arising from spacetime non-commutativity. Since this correction is positive, it partially compensates the negative pressure of the commutative case, thereby reducing its magnitude. It is worth emphasizing that the pressure correction obtained here differs from that reported in Ref.~\cite{25}, owing to the corrected event-horizon shift and the corresponding thermodynamic quantities derived in the present work.

\begin{figure}[H]
    \centering
    \includegraphics[width=0.9\linewidth]{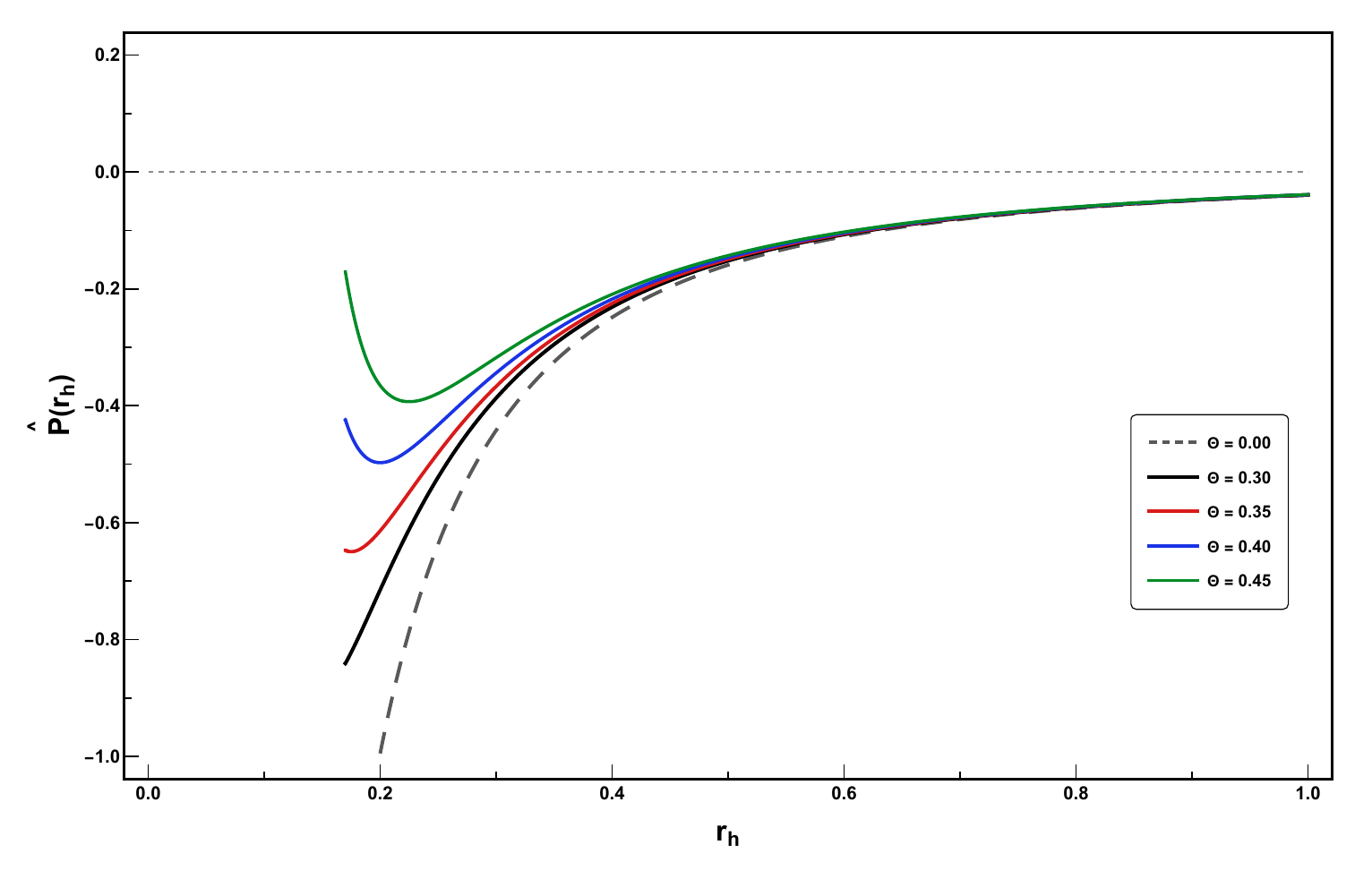}
    \caption{Corrected pressure $\hat{P}(r_h)$ of the non-commutative Schwarzschild black hole for different values of the non-commutative parameter $\Theta$. For $\Theta\neq0$, the pressure exhibits a minimum.}
    \label{fig:Phat_rh}
\end{figure}

\subsection{Evaporation process}

To estimate the lifetime of the non-commutative Schwarzschild black hole, we employ the Stefan--Boltzmann law, which describes the rate of energy loss due to Hawking radiation \cite{37},
\begin{equation}
\frac{dM}{dt}=-a\tilde{\alpha}\sigma T^4,\label{41}
\end{equation}
where $a$ is the radiation constant, $\tilde{\alpha}$ is the greybody factor, and $\sigma$ denotes the effective emission area. Using the geometric-optics approximation, $\sigma=27\pi M^2$, together with the corrected Hawking temperature, the mass-loss rate in non-commutative spacetime becomes
\begin{equation}
\frac{d\hat{M}(M,\Theta)}{dt}=-\frac{27\pi a\tilde{\alpha}}{\left(8\pi\right)^4}\hat{T}^2\left(M,\Theta\right).\label{42}
\end{equation}

Substituting the corrected expressions for the black-hole mass and Hawking temperature into Eq.~(\eqref{42}) yields the evolution equation
\begin{equation}
\left(1-\frac{\Theta^2}{16M^2}\right)\frac{dM}{dt}=-\frac{27a\tilde{\alpha}}{(8\pi)^4M^2}\left(1-\frac{9\Theta^2}{8M^2}\right).\label{43}
\end{equation}

Integrating Eq.~(\ref{43}) perturbatively up to second order in the non-commutative parameter $\Theta$, from the initial black-hole mass $M_{\mathrm{i}}$ to its mass $M$ at a later time $t$, gives
\begin{equation}
t=\frac{M_{\mathrm{i}}^3-M^3}{3\lambda}\left(1+\frac{17}{16M_{\text{i}}M}\Theta^2\right),\label{44}
\end{equation}
where
\begin{equation}
\lambda=\frac{27a\tilde{\alpha}}{\left( 8\pi \right) ^{4}},\label{45}        
\end{equation}
and $M_{\mathrm{i}}$ denotes the initial black-hole mass.

It is evident from Eq.~(\ref{44}) that the evaporation time increases significantly as the black-hole mass decreases. Within the perturbative regime considered here, the non-commutative corrections always increase the lifetime of the Schwarzschild black hole relative to its commutative counterpart. This behavior can be interpreted as a consequence of the effective contribution of the non-commutative parameter to the black-hole mass, which slows down the evaporation process. Moreover, the existence of a finite remnant mass suggests that black holes do not evaporate completely. Such remnants may constitute the primordial seeds of black holes, which could subsequently grow through accretion and mergers to form the massive black holes observed today. From this perspective, spacetime non-commutativity may play a fundamental role during the earliest stages of black-hole formation.

\section{Conclusion}

In this work, we have investigated the thermodynamic properties of the Schwarzschild black hole within the framework of non-commutative (NC) gauge gravity by employing the corrected non-commutative tetrad fields derived in Ref.~\cite{26}. Using the corrected metric, we obtained a modified expression for the event-horizon radius, which exhibits a qualitatively new behavior: the black hole possesses a minimum mass, $M^{\rm min}$, below which no event horizon exists. We also derived the corresponding non-commutative corrections to the Hawking temperature and showed that they remove the divergence present in the commutative Schwarzschild solution. As a consequence, the Hawking temperature reaches a finite maximum during the evaporation process before decreasing to zero at a minimum horizon radius $r_h^{0}$, corresponding to the minimum mass $M^{\rm min}$. At this stage, the Hawking radiation ceases, leaving behind a finite black-hole remnant. Furthermore, our analysis suggests that the characteristic scale of spacetime non-commutativity is of the order of the Planck length. In addition, we derived the non-commutative corrections to the entropy, heat capacity, Helmholtz free energy, and thermodynamic pressure.

The corrected entropy exhibits a logarithmic correction to the Bekenstein--Hawking area law and further supports the existence of a minimum remnant mass at which the evaporation process terminates. The analysis of the heat capacity reveals a rich thermodynamic phase structure consisting of three distinct regimes: two thermodynamically unstable phases with negative heat capacity, corresponding to small black holes ($0<r_h<r_h^{(1)}$) and large black holes ($r_h>r_h^c$), separated by an intermediate thermodynamically stable phase with positive heat capacity for $r_h^{(1)}<r_h<r_h^c$. Moreover, the divergence of the heat capacity identifies the critical point associated with a second-order phase transition. The behavior of the Helmholtz free energy is fully consistent with these stability properties and further confirms the existence of a thermodynamically preferred configuration.

Finally, we investigated the evaporation process of the non-commutative Schwarzschild black hole. We found that spacetime non-commutativity increases the evaporation time relative to the classical Schwarzschild case by introducing corrections that become significant at small horizon radii. These corrections effectively slow down the evaporation process and prevent the complete disappearance of the black hole, leading instead to the formation of a finite remnant. Such remnants may provide valuable insight into the final stage of black-hole evaporation and could play an important role in understanding the interplay between quantum gravity and black-hole thermodynamics. 

\section*{Acknowledgments}

This work is supported by PRFU research project B00L02UN050120230003, Univ. Batna 1, Algeria.

\end{document}